# S-curve networks and an approximate method for estimating degree distributions of complex networks[*]

Guo Jin-Li (郭进利)

*Business School, University of Shanghai for Science and Technology, Shanghai,*

*200093 ,China*

In the study of complex networks almost all theoretical models have the property of infinite growth, but the size of actual networks is finite. According to statistics from the China Internet IPv4 (Internet Protocol version 4) addresses, this paper proposes a forecasting model by using S curve (Logistic curve). The growing trend of IPv4 addresses in China is forecasted. There are some reference value for optimizing the distribution of IPv4 address resource and the development of IPv6. Based on the laws of IPv4 growth, that is, the bulk growth and the finitely growing limit, it proposes a finite network model with a bulk growth. The model is said to be an S-curve network. Analysis demonstrates that the analytic method based on uniform distributions (i.e., Barabási-Albert method) is not suitable for the network. It develops an approximate method to predict the growth dynamics of the individual nodes, and use this to calculate analytically the degree distribution and the scaling exponents. The analytical result agrees with the simulation well, obeying an approximately power-law form. This method can overcome a shortcoming of Barabási-Albert method commonly used in current network research.

[*] Project supported by the National Natural Science Foundation of China ( Grant No. 70871082), and the Shanghai Leading Academic Discipline Project under Grant No S30504.

## 1. Introduction

In the past decade, network science has made some breakthroughs. More recently network science efforts have focused on mathematically describing different network topologies. In 1998, Watts and Strogatz developed a small world network.[1,2] In 1999, Barabási and Albert reconciled empirical data on networks with mathematical representation, describing a scale-free network. Small world effect and scale-free characteristics in complex networks have a high degree of universality. This has aroused the concern and attention at home and abroad.

Since ten years ago, Barabási and Albert have made groundbreaking work on scale-free networks. Complex networks focus on empirical studies, synchronization of complex networks, evolving models of networks, and so on.[1,2-10] The empirical researches on complex networks report scale-free phenomena, what kind of model can depict the actually scale-free network are particularly important. Many scholars generalized and modified the BA (Barabási and Albert) model, such as models with nonlinear preferential attachment, with dynamic edge rewiring, fitness models and deterministically growing models can be found in the literature.[4] In real-life networks, incomers may only connect to a few others in a local area for their limited information, and individuals in a local area are likely to have close relations. Accordingly, we propose a local preferential attachment model.[8] Here, a local-area-network stands for a node and all its neighbors. Tanaka and Aoyagi present a weighted scale-free network model, in which the power-law exponents can be controlled by the model parameters.[9] The network is generated through the weight-driven preferential attachment of new nodes to existing nodes and the growth of the weights of existing links.[9] In sociology the preferential attachment is referred to as the Matthew effect. Price called it cumulative advantage.[6] When Price studied the citation network, in order to avoid a zero probability that a new node attaches to an isolate node he suggested that for each node attached an initial attractiveness.[6] One can consider the initial publication of a paper to be its first citation (of itself by itself).[6] Dorogovtsev *et. al.* further discussed the network with the initial attractiveness. They used the initial attractiveness to tune a power-law exponent. This attractiveness is a constant, and the attractiveness between nodes is independent of each other. Obviously, this is not reasonable. In the real world, many complex networks have a changing attractiveness over time. For example, WWW pages have the timeliness. Some new Web pages are fresh. Their attractiveness may be larger. As time goes by, the fresh may gradually disappear. The attractiveness of these Web pages may also gradually decrease. In the citation network, a paper is not the reference itself when it is published, but with an evolution of the citation network, its contents have a certain attraction. To describe this kind of networks, a competitive network with dynamic attractiveness is proposed in Refs.[11, 12], the fitness model[13] is a special case of the competitive network.

There are two common features of networks mentioned above. Nodes are linear growth,

that is, to add a single node at each time steps; the networks are infinitely growing. However, the growth of actual networks is limited. In the literature, except Refs.[7,14] almost all models grow to be unlimited. Are there any finitely growing networks in the real world? How can we establish a finitely growing model for complex networks? What is topology of this kind of networks?

Bulk arrival is important as well. Barabasi et al. proposed a deterministic model of geometrically growing networks.[15] In order to demonstrate that the metabolic networks have features such as modularity, locally high clustering coefficients and a scale-free topology, Ravasz et al. proposed a deterministic model of geometrically growing network.[16] Inspired by the idea of Ravasz *et al*., a random model of geometrically growing network is proposed.[17] It grows by continuously copying nodes and can better describe metabolic networks than the model of Ravasz et al. Concerning a problem of Apollonian packing, the geometrically growing network with common ratio 3 was introduced by Andrade et al. in 2005. Unfortunately, the degree distribution of Apollonian network is incorrect and the method using by Andrade et al. exits a regrettable shortcoming. We firstly found this error, and modified it.[17,18] The analytic method based on uniform distributions (i.e., Barabási-Albert method) is also not suitable for the analysis of the geometrically growing network.[17] In fact, Barabási-Albert method have a shortcoming for the analysis of other models as well.[19] Up to now, the finite networks with bulk growth are studied to be relative rarity. This paper aims to propose a finite model with bulk growth. We develop a method to predict the growth dynamics of the individual nodes.

The paper is organized as follows. In Sec. 2 we collected the China's IPv4 (Internet Protocol version 4) data. It is fitted by Logistic curve (S curve). We forecast the growing trend of China's IPv4. In Sec. 3 we address above some questions. According to a law of IPv4 growth, we propose a finite network model with bulk growth. The model is called S-curve network. In Sec. 4 we firstly show that the Barabási-Albert method is not suitable for the analysis of the model. We develop the method to predict the degree distribution of complex networks, and estimate the degree distribution of S-curve network. In Sec. 5 simulations of the network are given. The analytical result agrees with the simulation well, obeying an approximately power-law form. Conclusions are drawn in Sec. 6.

## 2. Study on IPv4 standard diffusion tendency in China

### 2.1 Data set

IPv4 is the fourth revision in the development of the Internet Protocol (IP) and IPv6 (Internet Protocol version 6) is the sixth revision. IPv4 is the first version of the protocol to be widely deployed. Together with IPv6, it is at the core of standards-based internetworking methods of the Internet. IPv4 is still by far the most widely deployed Internet Layer protocol. Using this protocol, people find that Internet implements interconnection between different hardware structures and different operating systems. In Internet, each node based on the unique IP addresses ties each other. IPv4 uses only 32 bits. Internet Protocol version 4 provides approximately 4.3 billion addresses. China Internet Network Information Center annually publishes "Statistical Survey Report on the Internet Development in China". 25th statistical survey report on the internet development in China pointed out that by December 2009, the number of Chinese IPv4 addresses had reached 232 million, up 28.2% from late 2008. Recent two years, the average IPv4 addresses of Internet users fell slightly; meanwhile, IPv4 addresses are facing the exhaustion. This data is obtained from the previous released by China Internet Network Information Center "Statistical Survey Report on the Internet Development in China". Table 1 is the number of China IPv4 addresses over time.

**Table 1.** Accumulative number of Chinese IPv4 address

| Time | n(IPv4)/$10^4$ | Time | n(IPv4)/$10^4$ |
|---|---|---|---|
| 2002-12 | 2900 | 2006-12 | 9802 |
| 2003-06 | 3208 | 2007-06 | 11825 |
| 2003-12 | 4146 | 2007-12 | 13527 |
| 2004-06 | 4942 | 2008-06 | 15814 |
| 2004-12 | 5995 | 2008-12 | 18127 |
| 2005-06 | 6830 | 2009-06 | 20503 |
| 2005-12 | 7439 | 2009-12 | 23245 |
| 2006-06 | 8479 | | |

## 2.2 Forecasting model

Logistic curve model, it is also called S curve. It is developed by biological mathematicians Verhulst in 1845 to study population growth. Unfortunately, it was neglected for a long time, Pearl and Reed rediscovered its applications in the 1920s of the 20th century. Logistic curve are widely used in economics, politics, population statistics, human tumor proliferation, chemistry, plant population dynamics, insect ecology and forest growth.

Logistic curve

$$N = \frac{L}{1 + e^{a-rt}} \qquad (1)$$

Where $N$ is the growth; $t$ is time; $r$ is a constant, it is called an instantaneous growth rate; $L$ is also the constant, it is called carrying capacity. Coefficient $a$ determines the position of Logistic curve on time axis. If we find $a$, $r$ and $L$, the curve is obtained. Based on four points method[20] and data in table 1, we have $L$=140359.5

According to the regression analysis, we obtain

$$r \approx 0.159, \quad a \approx 4$$

Thus, Logistic fitting curve is as follows

$$N = \frac{140359.5}{1 + 54.6e^{-0.159t}} \qquad (2)$$

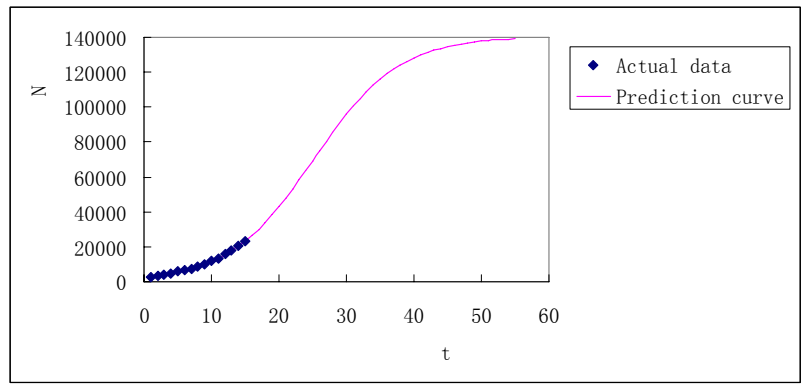

**Fig.1** Logistic curve fitting the actual number of IPv4

**Table 2.** Accumulative number of Chinese IPv4 address comparing with the prediction of Logistic curve

| Time | n(IPv4)/10$^4$ | Prediction n(IPv4)/10$^4$ | Percentage of error |
|---|---|---|---|
| 2002-12 | 2900 | 2950.365254 | 1.736732906 |
| 2003-06 | 3208 | 3446.340589 | 7.414568936 |
| 2003-12 | 4146 | 4023.251346 | -2.960652543 |
| 2004-06 | 4942 | 4693.425271 | -5.029840735 |
| 2004-12 | 5995 | 5470.754013 | -8.74472038 |
| 2005-06 | 6830 | 6370.778927 | -6.723588192 |
| 2005-12 | 7439 | 7410.737253 | -0.379926692 |
| 2006-06 | 8479 | 8609.548973 | 1.53967417 |
| 2006-12 | 9802 | 9987.719943 | 1.894714786 |
| 200706 | 11825 | 11567.13249 | -2.180697731 |
| 2007-12 | 13527 | 13370.69111 | -1.155532567 |
| 2008-06 | 15814 | 15421.78953 | -2.480147132 |
| 2008-12 | 18127 | 17743.56799 | -2.115253548 |
| 2009-06 | 20503 | 20357.93743 | -0.707518768 |
| 2009-12 | 23245 | 23284.36345 | 0.169341574 |
| 2010-06 |  | 26538.42785 |  |
| 2010-12 |  | 30130.22053 |  |
| 2011-06 |  | 34062.65756 |  |
| 2011-12 |  | 38329.86696 |  |
| 2012-06 |  | 42915.82491 |  |
| 2012-12 |  | 47793.44875 |  |

## 3. Description of the model

### 3.1. Competitive network model

Ten years ago, the mechanism of cumulative advantage proposed by Barabási and Albert is now widely accepted as the probable explanation for the power-law degree distribution observed not only in WWW network but in a wide variety of other networks also. In fact, for the mechanism of citation networks has been proposed by Price three decades ago.[6,21] The work of Price himself, however, is largely unknown in the scientific community, and cumulative advantage did not achieve currency as a model of network growth until its rediscovery some decades later by Barabási and Albert.[6]

The BA model satisfies following rules. Starting with a small number（$m_0$）of nodes, at every time step, we add a new node with $m\ (\leq m_0)$ edges that link the new node to $m$ different nodes already present in the system; When choosing the nodes to which the new node connects, we assume that the probability $\Pi$ that a new will be connected to node $i$ depends on the degree $k_i$ of node $i$, such that $\Pi(k_i) = \dfrac{k_i}{\sum\limits_{j} k_j}$. In order to avoid the isolated node is never connected, initial attractiveness $A$ of nodes is introduced.[10] Because $A$ is a constant, the model in Ref.[10] can not reflect the competition of networks. To describe the competition, we proposed Poisson NPA competition model. The algorithm of our model is as follows. (1) *Random growth:* Starting with a small number （$m_0$）of nodes. The arrival process of nodes is a Poisson process having rate $\lambda$. At time t, if a new node is added to the system, the new node is connected to $m\ (\leq m_0)$ different nodes already present in the system; (2) *Preferential attachment:* When choosing the nodes to which the new node connects, we assume that the probability $\Pi$ that a new will be connected to node $i$ depends on the degree $k_i(t)$ of node $i$ and its attractiveness $a_i(t)$, such that $\prod(k_i(t)) = \dfrac{k_i(t) + a_i(t)}{\sum\limits_i (k_i(t) + a_i(t))}$, where, at time $t$, attractiveness $a_i(t)$ of node $i$ and the degree of nodes of the network is independent of each other, or $\dfrac{a_i(t)}{k_i(t)}$ is chosen from a distribution $F(x)$; $N(t)$ denotes the number of nodes at time $t$; and

$$a_i(t) > 1 - k_i(t), i = 1, 2, \cdots, \quad \sum_{i=0}^{N(t)} a_i(t) = O(t^\theta), \quad 0 \le \theta \le 1.$$

### 3.2. S-curve network

An IPv4 is as a node and the contact between two IPv4 as an edge. If there is any contact between two IPv4, one edge is added between the two nodes. Therefore, all IPv4 in China form a complex network. It is called as an IPv4 network. Its key features are as follows. (1) Nodes are bulk growth; (2) the limit of the growth is finite, accumulative number of nodes is Logistic curve. Based on the two features, we propose an S-curve network to be as follows. (a) *S-curve growth*: Starting with a small number（ $m_0$ ）of nodes. At time t, we add $\dfrac{rL\exp(a - rt)}{(1 + \exp(a - rt))^2}$ new nodes to the system. Each new node is individually connected to $m$ $(\le m_0)$ different nodes already present in the system, and all the new nodes do not connected to each other; (b) *Preferential attachment:* When choosing the nodes to which the new node connects, we assume that the probability $\Pi$ that a new will be connected to node $i$ depends on the degree $k_i$ of node $i$, such that

$$\prod(k_i(t)) = \frac{k_i(t) + A}{\sum_i (k_i(t) + A)}, \tag{3}$$

where $A$ is the initial attractiveness of node $i$, and $A > -m$.

## 4. Analysis of the model

### 4.1 Estimation of network size

For convenience, taking $m_0 \approx \dfrac{L}{1 + \exp(a)}$, at $t = 0$, first batch nodes are added to the system. Then at time $t$ the number of nodes of the network is as follows

$$N(t) = m_0 + \int_0^t \frac{rL\exp(a - rt)}{(1 + \exp(a - rt))^2}\,dt = m_0 + \frac{L}{1 + \exp(a - rt)} - \frac{L}{1 + \exp(a)} = \frac{L}{1 + \exp(a - rt)}$$

$$\tag{4}$$

$$\lim_{t \to \infty} N(t) = \lim_{t \to \infty} \frac{L}{1 + \exp(a - rt)} = L \tag{5}$$

Therefore, although the network is growing, the network size is finite. It displays S curve, see Fig. 2.

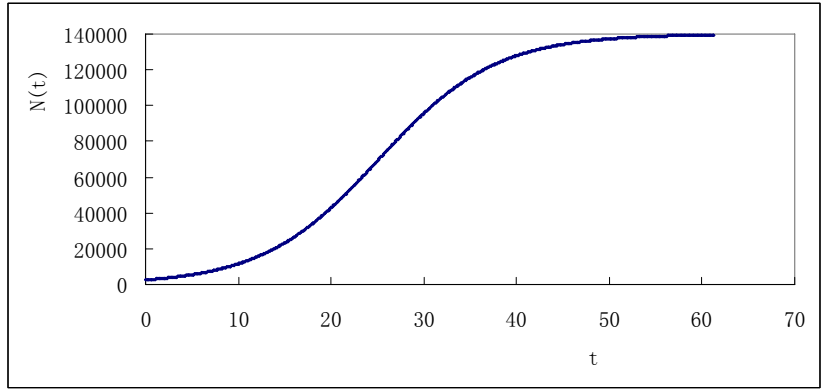

**Fig. 2** Schematic diagram of the size of S-curve network

## 4.2 Degree distribution based on Barabási-Albert method

$t_i$ denotes the time that the $i$th batch nodes is added to the system. For convenience, let us label each node of the $i$th batch nodes with $j$ accordingly. $k_{ij}$ denotes the degree of the $j$th node in the $i$th batch at time $t$. Assuming that $k_{ij}(t)$ is a continuous real variable, the rate at which $k_{ij}(t)$ changes is expected to be proportional to degree $k_{ij}(t)$ of the $j$th node in the $i$th batch. Consequently $k_{ij}(t)$ satisfies the dynamical equation

$$\frac{\partial k_{ij}}{\partial t} = m \frac{rL \exp(a - rt)}{(1 + \exp(a - rt))^2} \frac{k_{ij} + A}{\sum_{i,j}(k_{ij} + A)}. \tag{6}$$

Taking into account

$$\sum_{i,j}(k_{ij} + A) = 2mN(t) + AN(t) = \frac{mL}{1 + \exp(a - rt)}(2 + \frac{A}{m})$$

Substituting this into Eq. (6) we obtain

$$\frac{\partial k_{ij}}{\partial t} = \frac{r \exp(a - rt)}{(2 + \frac{A}{m})(1 + \exp(a - rt))}(k_{ij} + A), \tag{7}$$

The solution of this equation, with the initial condition $k_{ij}(t_i) = m$, is

$$k_{ij}(t) = (m+A)\left(\frac{1+\exp(a-rt_i)}{1+\exp(a-rt)}\right)^{\frac{1}{2+\frac{A}{m}}} - A \ .$$  （8）

Using Eq. (8), one can write the probability that a node a degree $k_{ij}(t)$ smaller than $k$, $P\{k_{ij}(t) < k\}$, as

$$P\{k_{ij}(t) < k\} = P\{t_i > \frac{a}{r} - \frac{1}{r}\ln[(\frac{k+A}{m+A})^{2+\frac{A}{m}}(1+\exp(a-rt))-1]\}$$  (9)

Assuming that we add the batch of nodes at equal time intervals to the network, the $t_i$ values have a constant probability[2,3]

$$P(t_i) = \frac{1}{m_0 + t}$$  (10)

Substituting this into Eq. (9) we obtain

$$P\{k_{ij}(t) < k\} = 1 - \frac{1}{t+m_0}[\frac{a}{r} - \frac{1}{r}\ln[(\frac{k+A}{m+A})^{2+\frac{A}{m}}(1+\exp(a-rt))-1]]$$  (11)

The degree distribution $P(k)$ can be obtain using

$$P(k) = \frac{\partial P\{k_{ij}(t) < k\}}{\partial k} = \frac{(2m+A)(k+A)^{1+\frac{A}{m}}(1+\exp(a-rt))}{rm(t+m_0)((k+A)^{2+\frac{A}{m}}(1+\exp(a-rt))-(m+A)^{2+\frac{A}{m}})} \ ,$$

predicting that asymptotically ($t \to \infty$)

$$P(k) = 0, \qquad k = m+1, m+2, \cdots.$$  (12)

Eq. (12) shows that the analytic method based on uniform distributions (i.e., Barabási-Albert method) is not suitable for the model.

### 4.3. Approximate method for estimating distributions

For any given $i, h, t > 0$, and $h < i$, $t_h < t_i \leq t$, from (8), we can know

$$k_{hj}(t) = (m+A)\left\{\left(\frac{1+\exp(a-rt_h)}{1+\exp(a-rt)}\right)^{\frac{1}{2+\frac{A}{m}}} - A\right\} > (m+A)\left\{\left(\frac{1+\exp(a-rt_i)}{1+\exp(a-rt)}\right)^{\frac{1}{2+\frac{A}{m}}} - A\right\} = k_{ij}(t) \quad (13)$$

From (13) we know that the degree of the node entered the network before $i$th batch is almost everywhere greater than $k_{ij}$. Because the degree of the network may be continuous growth, that is, at time $t$ when $i$th batch nodes add to the system the degree of the network may be as follows, $m, m+1, m+2, \cdots, k_{ij}(t), \cdots, k_{0j}$. Thus, for any given integer $k_{0j}(t) \geq k > m$, there may exist $k_{ij}(t)$, such that, $k_{ij}(t) = k$. The number of nodes that the degree is greater than or equal to $k$ is $N(t_i)$. Thus, the cumulative degree distribution is obtained as follows

$$P_{cum}(k) = \frac{N(t_i)}{N(t)} = \left(\frac{m+A}{k_{ij}+A}\right)^{2+\frac{A}{m}} = \left(\frac{m+A}{k+A}\right)^{2+\frac{A}{m}} \quad (14)$$

which is the probability that the degree is greater than or equal to $k$. The cumulative distribution reduces the noise in the tail. Since the degree of the network may be continuous growth, from (14) we know that the network is scale-free, and degree exponent $\gamma = 3 + \frac{A}{m}$.

### 5. Simulation result

In succession, we show some simulations. The number of nodes of our simulation is $L = 120060$. Other parameters is as follows, $m_0 = 60$, $m = 8$, $a = 7$ and $A = 0$. The analytical result agrees with the simulation well, obeying an approximately power-law form with exponent 3, see Fig.3—Fig.6.

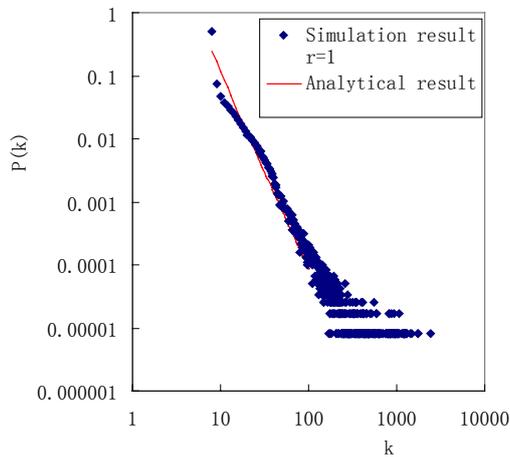

**Fig. 3.** The degree distribution of the present model with *r* = 1. The squares and solid curve represent the simulation and analytical results, respectively.

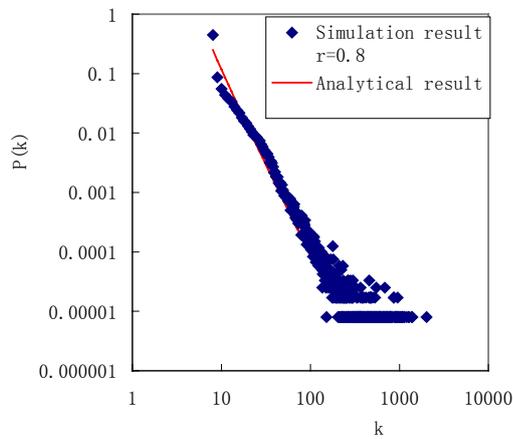

**Fig. 4**. The degree distribution of the present model with *r* = 0.8. The squares and analytical curve represent the simulation and analytical results, respectively.

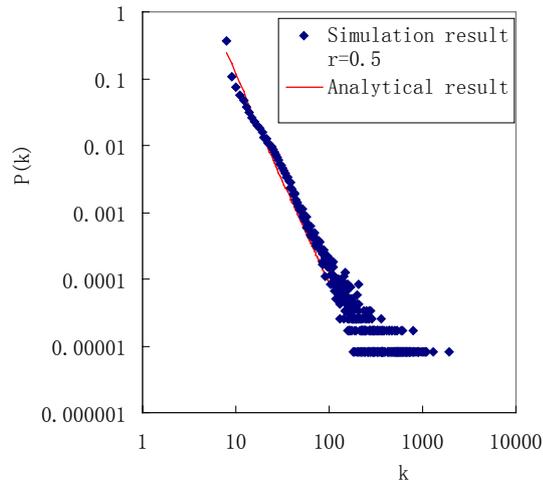

**Fig. 5**. The degree distribution of the present model with *r*= 0.5. The squares and solid curve represent the simulation and analytical results, respectively.

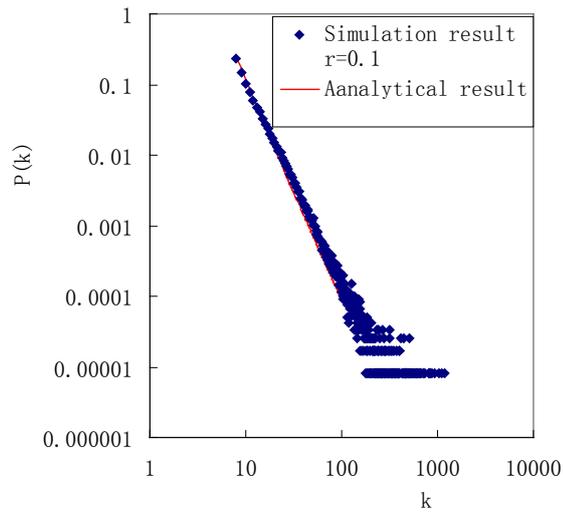

**Fig. 6**. The degree distribution of the present model with *r*= 0.1. The squares and solid curve represent the simulation and analytical results, respectively.

## 6. Summary and conclusions

Using Logistic model we study IPv4 standard diffusion tendency in China. We forecast IPv4 diffusion tendency. Our result shows that the limited resources of IPv4 will be a great restriction on china's Internet development. Therefore, we suggest that the country should step up IPv6 standards related to the development and marketing efforts. From the strategic high, the development of the next-generation Internet is a task of top priority.[22]

The network formed by IPv4 is a bulk growth and the limit of the growth is finite. The accumulative number of the network is S curve. A lot of practical networks have S-curve characteristics such as the network formed by Internet users. We propose the S-curve network model based on the feature of IPv4 network. Barabási and Albert assumed that the time that the node is added to the system obeys uniform distributions. Although this assumption is not reasonable, it does not affect the result of estimating the distribution of singly growing networks.[2] However, Barabási-Albert method can not estimate the degree distribution of bulk growing networks. To overcome the difficulty, we propose an approximate method for estimating the degree distribution of complex networks. Simulation results show that this method is effective. Although this is not a rigorous method, it is an intuitive and effective method for estimating the degree distribution. The rigorous method for analyzing complex networks needs a stochastic process theory.[7,12,19]